\documentclass{aa}  
\usepackage{graphicx}
\usepackage{txfonts}

\usepackage{natbib}
\bibpunct{(}{)}{;}{a}{}{,}

\def\kms {{\mathrm{km}\,\mathrm{s}^{-1}}}

\def\msol {{\mathrm{M}_\odot}}
\def\sun {{_{\odot}}}
\def\rsol {{R\sun}}
\def\cd  {{$\mbox{c~d}^{-1}$}}

\begin{document}
  \title{Multiperiodic pulsations in the Be stars NW~Ser and V1446~Aql\thanks{
Table A.1 and A.2 are only available in electronic form at the CDS via anonymous 
ftp to cdsarc.u-strasbg.fr (130.79.128.5) 
or via http://cdsweb.u-strasbg.fr/cgi-bin/qcat?J/A+A/
}}

   \author{J. Guti\'errez-Soto\inst{1,5}, J. Fabregat\inst{1},
     J. Suso\inst{2},
     J. C. Su\'arez\inst{3}
     A. Moya\inst{3,4},  R. Garrido\inst{3}, \\
     A.-M. Hubert\inst{5}, M. Floquet\inst{5},
     C. Neiner\inst{5} \and Y. Fr\'emat\inst{6}
   }

   \offprints{juan.gutierrez-soto@uv.es}

   \institute{Observatorio Astron\'{o}mico. Universidad de Valencia.
     Edificio Institutos de investigaci\'{o}n. Pol\'{\i}gono la Coma,
     46980 Paterna, Valencia, Spain
     \and
     ICMUV, Edificio Institutos de investigaci\'{o}n. Pol\'{\i}gono la Coma,
     46980 Paterna Valencia, Spain
     \and  Instituto de Astrof\'{\i}sica de Andaluc\'{\i}a (CSIC)
     Camino Bajo de Hu\'{e}tor, 24, 18008 Granada, Spain
     \and LESIA, UMR 8109 du CNRS, Observatoire de Paris-Meudon, 92195 Meudon, France
     \and GEPI, UMR 8111 du CNRS, Observatoire de Paris-Meudon,  92195 Meudon, France
     \and Royal Observatory of Belgium, 3 Avenue Circulaire, B-1180 Brussels, Belgium
  }

   \authorrunning{J. Guti\'errez-Soto et al.}
   \titlerunning{Multiperiodic pulsations in NW Ser and V1446 Aql}

   \date{Received ; accepted }

  \abstract
   {}
   {We present accurate photometric time series of two Be stars: 
\object{NW Ser} and \object{V1446 Aql}. Both stars were observed at the 
Observatorio de Sierra Nevada (Granada) in July 2003 with an automatic 
four-channel Str\"omgren photometer. We also present a preliminary theoretical
 study showing that the periodic variations exhibited by these stars can be 
due to pulsation.}
   {An exhaustive Fourier analysis together with a least-square fitting has 
been carried out on the time series for all four Str\"omgren bands. 
Several independent frequencies and non-periodic trends explain most of the 
variance. A theoretical non-adiabatic code applied to stellar models for 
these stars shows that g-modes are unstable.}
   {Both stars show rapid variations in amplitude, probably due to a beating 
phenomenon. Four significant frequencies have been detected for each star.
Comparison of the observed amplitude ratios for each pulsational frequency 
with those calculated from theoretical pulsation codes allows us to estimate the 
pulsation modes associated with the different detected frequencies.
NW~Ser seems also to show unstable p-modes and 
thus could be one of the newly discovered $\beta$ Cephei and SPB hybrid stars.
Further spectroscopic observations are planned to study the 
stability of the detected frequencies.}
{}


   \maketitle
%

\section{Introduction}

Short-term variability, with periods ranging from a few hours to two 
days, is present in most of the early-type Be stars and in a significant 
fraction of late Be stars. It has been detected both spectroscopically, in the 
form of line profile variability ($lpv$) and photometrically, by means 
of non-regular light curves.

\citet{baade82} attributed the short periodic $lpv$ to non-radial 
pulsations ($nrp$). Further detailed studies of a few well-observed 
objects led to a complete and consistent multimode modelling of the 
observed variations \citep{rivinius01,maintz03,neiner05}. 
In addition, high-precision photometric data obtained with the MOST satellite showed the presence of 
multiple periods in the Be stars $\zeta$-Oph \citep{walker05b} 
and HD 163\,868 \citep{walker05a}.
They modelled the detected frequencies as $nrp$ in terms of p-modes in the case of $\zeta$-Oph, and g- and r-modes in the case of  HD 163\,868.

Similarly, COROT will provide important clues to understand the link between the beating of $nrp$ modes and the origin of the Be phenomenon.
COROT is based on ultra high precision, wide field photometry for very long continuous observing runs in the same field of view \citep{baglin02}. 
In preparation for this mission, a total of 84 Be stars have been studied photometrically in order to select the most suitable candidates 
to be observed by COROT (Guti\'errez-Soto et al., in preparation). 
Here we present the analysis of the photometric time series for two Be stars, namely NW~Ser and V1446~Aql.

NW~Ser (HR 6873, HD 168797, V=6.14) is a bright and extensively observed 
B2.5IIIe star, which has been photometrically monitored for over two decades.
 From the analysis of the Hipparcos photometric data, several authors have 
obtained similar results: \citet{hubert98} found short-term variability with 
a period of 0.488 days; \citet{percy99} reanalysed the Hipparcos data together
 with ground-based photometry finding a similar period of 0.46 days, as well
 as a longer period of 5.5 days; and finally, two periods (0.475 and 0.406 
days) have been detected by \citet{aerts00} using only Hipparcos data, 
although the author claimed that this finding should be confirmed by means
 of ground-based observations. 

V1446~Aql (HD 179405, B2IVe) has been observed to be an emission-line star in the 
Mount Wilson objective prism all-sky survey  \citep{merril43}. It appears as 
an irregular variable star from the Hipparcos data, which led to its inclusion 
in the GCVS \citep{kazarovets99}, although no short-term periodic variability 
had been detected so far. 

\begin{table}
\caption{Comparison and check stars used in the differential photometry.}
\centering
\begin{tabular}{ cccc }
\hline
\hline
  &   Star &  V &  Spectral Type \\
\hline
variable & NW~Ser &  6.14 & B2.5IIIe \\
comparison &  \object{HD 170\,200} & 5.71 & B8III-IV \\
check & \object{SAO 123\,607} & 8.6 & B8 \\
\hline
variable & V1446~Aql & 9.12 & B2IVe \\
comparison & \object{HD 179\,846} & 8.29 & B8 \\
check & \object{HD 178\,598} & 9.45 & B8 \\ 
\hline
\end{tabular}
\label{tabla comp}
\end{table}

\section{Observations and frequency analysis}

\begin{figure}
   \centering
   \includegraphics[width=7cm]{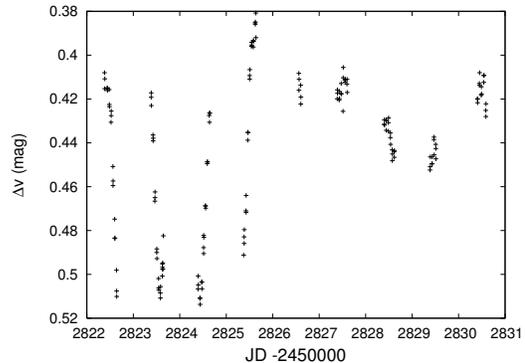} 
  \caption{Light curve of NW~Ser in the $v$ filter.}
   \label{fig curva nwser}
   \end{figure}

Observations were made with the 0.9 m telescope of the Observatorio de 
Sierra Nevada (OSN, Granada, Spain) between July 1 and 9 in 2003.
The instrument used is an automatic four-channel photometer, which allows 
simultaneous observations through the four $uvby$ filters of the Str\"omgren 
photometric system. The data discussed here and presented in Tables A.1 and A.2
 are the differential magnitude in 
the instrumental system between the variable and the comparison star of 
Table~\ref{tabla comp}. A check star has also been observed to verify 
possible intrinsic variations of the comparison star. Data have been corrected for 
sky background and atmospheric extinction. Table~\ref{tabla precision} shows 
the photometric precision as given by the standard deviation of
the difference between the comparison and check stars for the whole campaign for each filter.
The light curves of the stars are depicted in Figs.~\ref{fig curva nwser} 
and~\ref{fig curva v1446aql}, only the $v$ filter data are shown for clarity.

\begin{table}
\caption{Photometric precision in the instrumental system.}
\centering
\begin{tabular}{ ccccc }
\hline
\hline
Star  &   u filter &  v filter & b filter & y filter \\
\hline
NW~Ser& 0.010 & 0.005  & 0.005 & 0.005\\
V1446~Aql&0.011 & 0.007 & 0.007 & 0.008\\
\hline
\end{tabular}
\label{tabla precision}
\end{table}

The period analysis has been performed by means of standard Fourier analysis and 
least-square fitting. We have used Period04 \citep{lenz05} which is 
especially designed for the analysis of time series containing gaps. This 
program finds the frequencies one by one by computing the Fourier Transform 
and then adjusts the parameters of a sinusoidal function using a 
least-square fitting. This frequency is then removed and a new step is 
started finding a new frequency, the subsequent least-square fitting is performed 
allowing the two frequencies to move in order to obtain the minimum variance.
The method is iterative and stops when removing of a new frequency is not 
statistically significant.

We also used a non-linear multi-parameter fitting code which scans a 
wide range in frequency based on \citet{vanicek71} and is explained in detail in 
\citet{zerbi97}. This code is also well-suited for our case, for which daily 
aliases are present in the periodogram, due to the fact that our observations 
have been obtained at only one site.  

The way in which we determine whether the frequencies are statistically 
significant or not is described in \citet{breger93} and basically consists 
of the calculation of the \emph{signal to noise ratio} (SNR), the noise being the 
average amplitude, within a 5 \cd\ frequency interval, of the residual 
periodogram after the prewhitening as previously 
explained around a frequency and the signal being the corresponding 
amplitude of that frequency. \citet{breger93} showed that this value must be 
greater than 4.

Frequencies, amplitudes and phases obtained for NW~Ser and V1446~Aql are 
presented in Tables~\ref{tabla freq nwser} and~\ref{tabla freq v1446aql} 
respectively for the four Str\"omgren bands. We also show the total fraction 
of the variance removed from the signal (R), the SNR and 
the $\sigma$ of the final residual $\sigma_{res}$. In 
Figs.~\ref{fig freq nwser} and~\ref{fig freq v1446aql} we display the 
successive periodograms and the spectral window for NW~Ser and V1446~Aql 
respectively, results are shown only for the $v$ filter data for clarity. 

Following \citet{montgomery99}, the expected error in frequency for 
uncorrelated observations can be derived from the equation 

\begin{displaymath}
 \sigma_{F} = \frac{\sqrt{6}}{\pi} \cdot \frac{\sigma_{n}}{A \cdot \sqrt{N} \cdot T}
\end{displaymath}

where $\mathrm{A}/\sigma_{n}$ indicates the SNR, N is the 
number of observations and T the time elapsed between the first and the last 
data point. As noted by \citet{czerny91}, correlations in the residuals of the 
fitting have to be taken into account multiplying the error frequency by 
$\sqrt{D}$, where D is the correlation length. D can be estimated by 
performing an autocorrelation analysis of the final residuals. In our case
the correlation length is 4. Therefore, we derive a formal error in frequency 
of $6-20 \times 10^{-4}$ \cd\ on frequencies F1 to F4 respectively for NW~Ser 
and $8-16 \times 10^{-4}$ \cd\ for V1446~Aql.

\subsection{NW~Ser}

\begin{figure*}

    \centering
    \includegraphics[width=5.5cm]{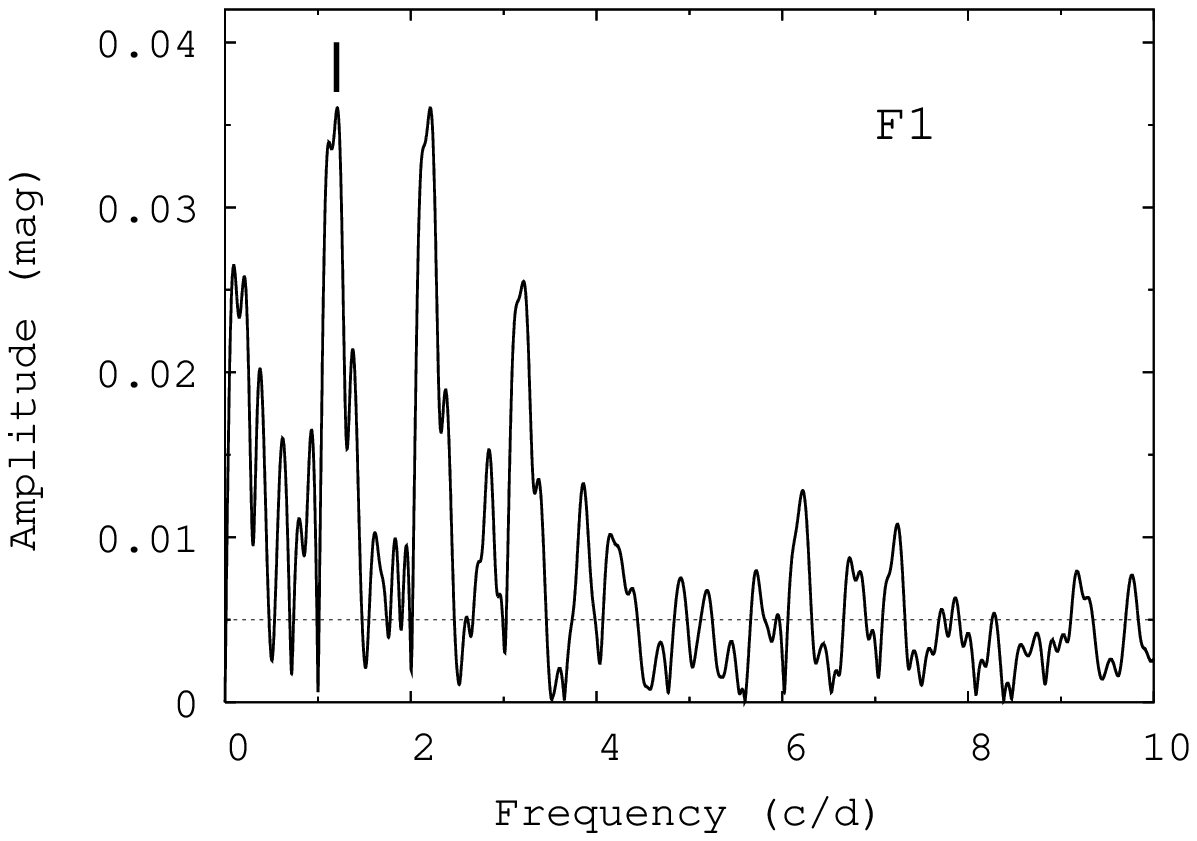}
    \includegraphics[width=5.5cm]{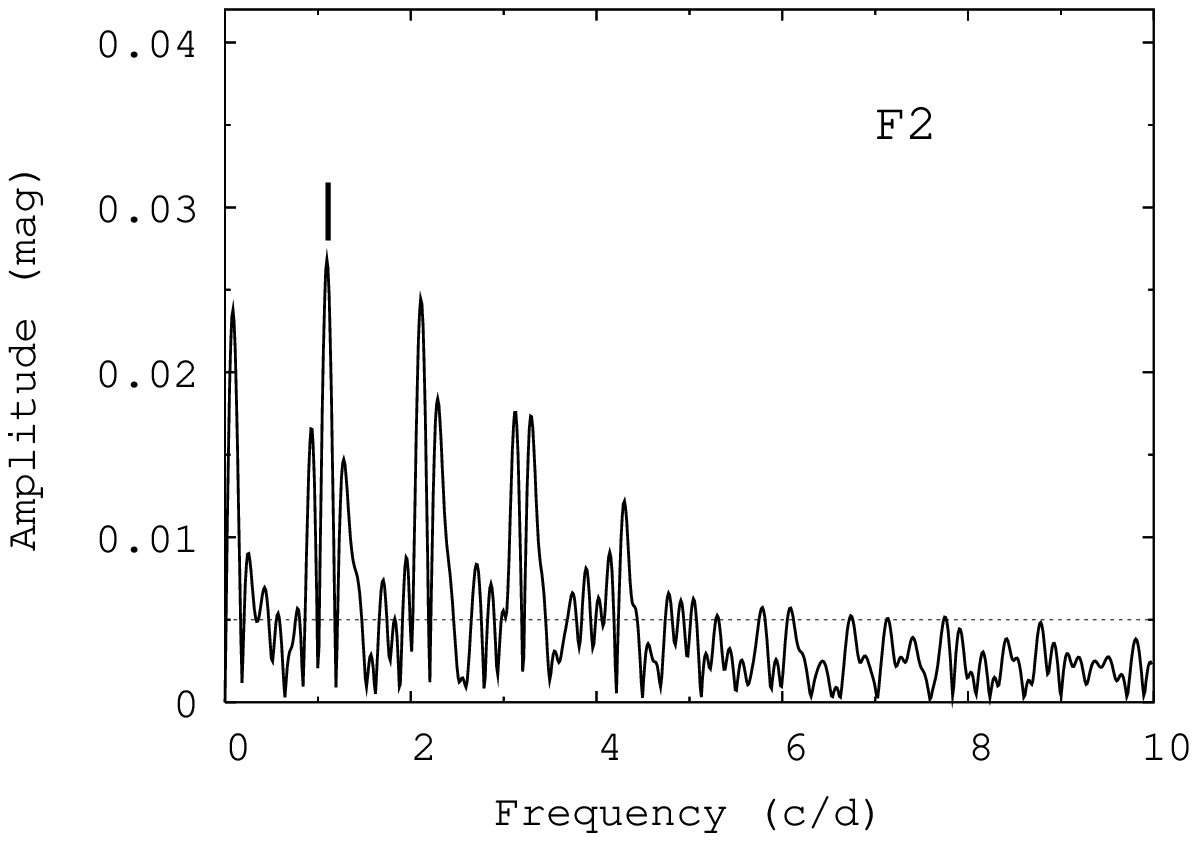}
    \includegraphics[width=5.5cm]{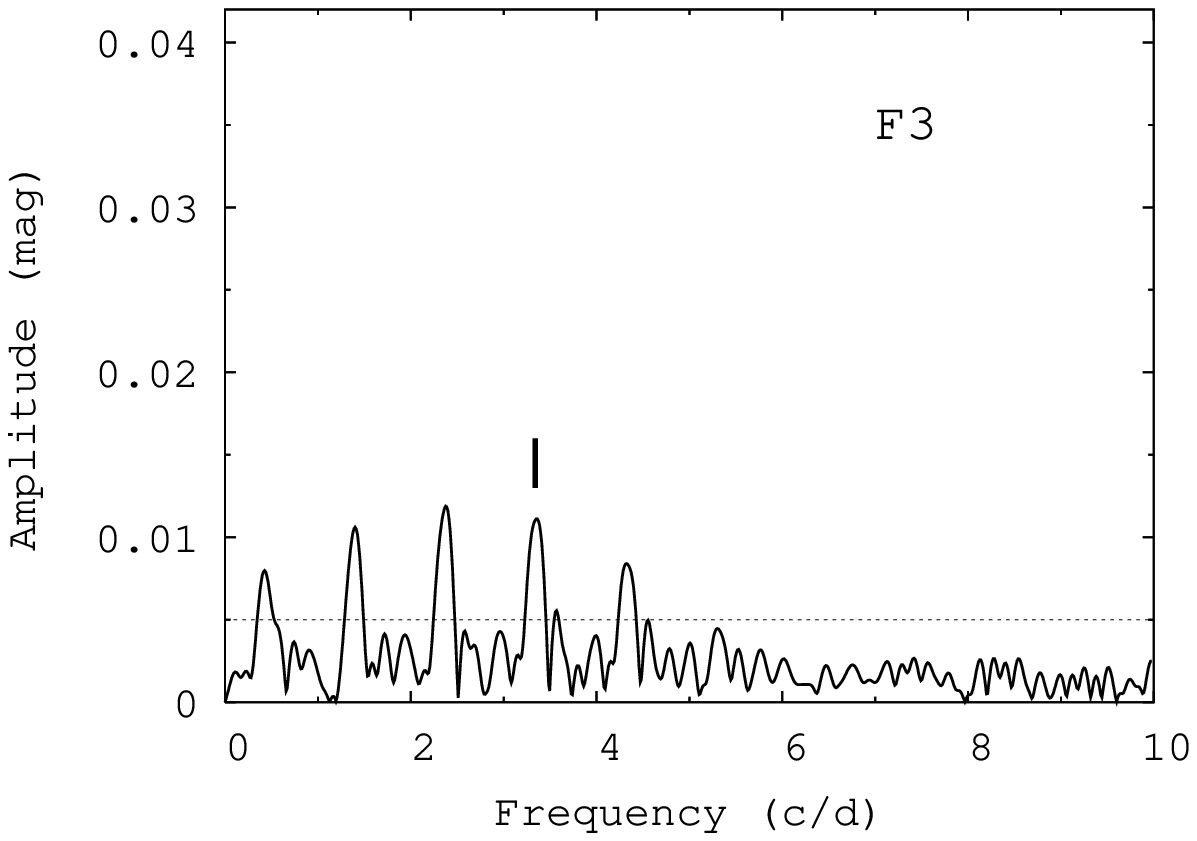}
    \includegraphics[width=5.5cm]{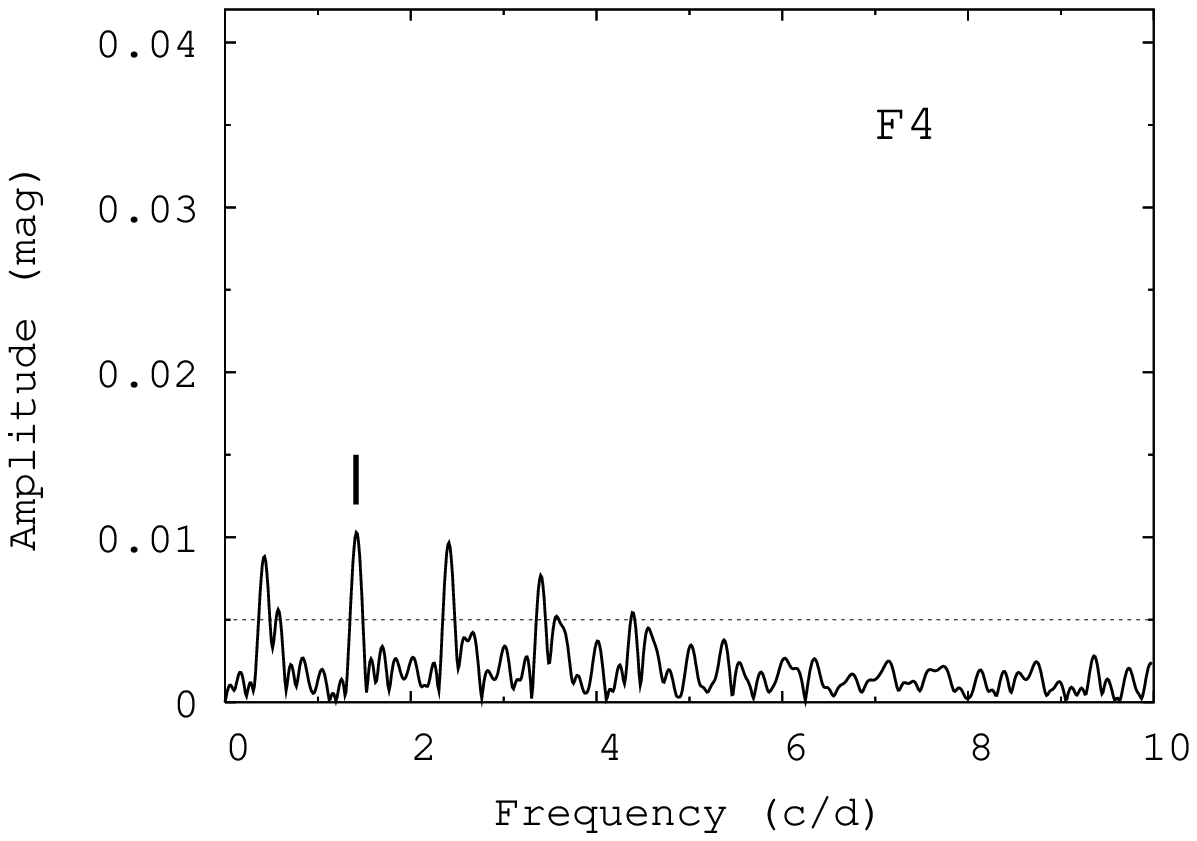}
    \includegraphics[width=5.5cm]{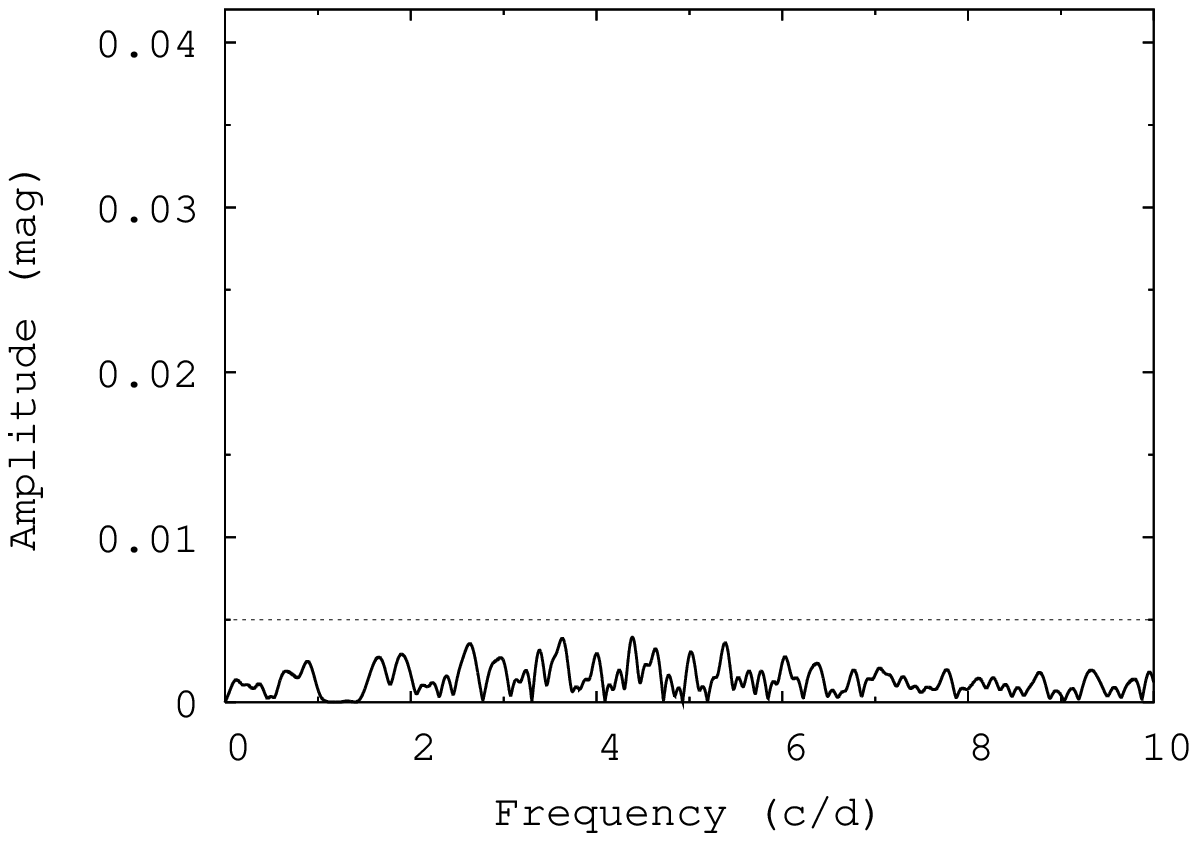}
    \includegraphics[width=5.5cm]{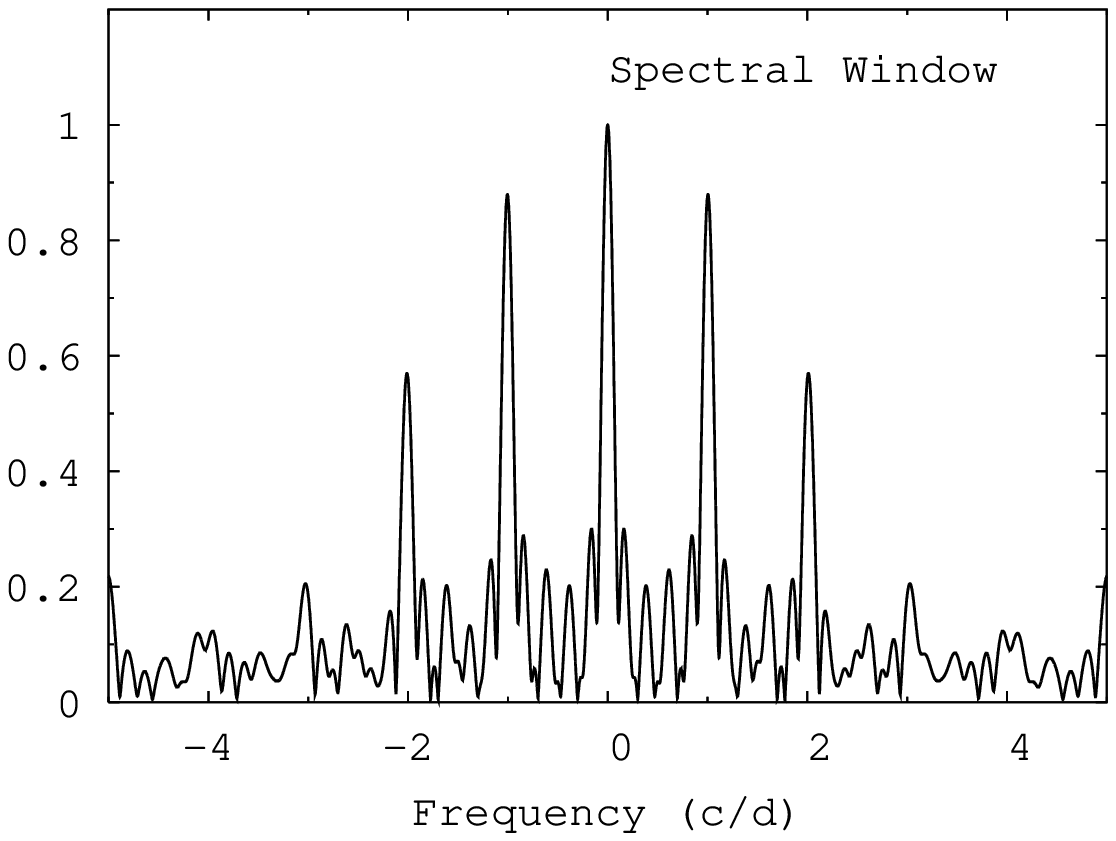}
    \caption{Successive periodograms for NW~Ser for the $v$ filter. The ticks 
indicate the positions of the detected frequencies and the dashed 
horizontal line indicates the 4$\sigma$ level after the final prewhitening.}
\label{fig freq nwser}
\end{figure*}

The rapid amplitude variation over the short timescale spanned by our 
observations strongly suggests the presence of a 
beating phenomenon. This would explain why after prewhitening for the first peak 
(seen at $\mathrm{F1}=1.197$ \cd\ in Fig.~\ref{fig freq nwser}, together with 
its aliases at 1,2,... \cd) a second peak appears at a very close position  
corresponding to a frequency at $\mathrm{F2}=1.126$ \cd (together with its 
daily aliases). Note that the width of the sidelobe is almost twice that 
calculated for the spectral window. This would mean that another frequency 
close to 1.197 \cd\ is also present. Following \citet{loumos78}, this occurs when 
the true frequencies differ by less than the half-width of the sidelobe and, in 
our case, a peak at the midpoint of the true frequencies would appear in the 
periodogram, as seen in Fig.~\ref{fig freq nwser}. F1 and F2 have 
comparable amplitudes, which would produce the rapid amplitude variability in 
terms of a beating phenomenon.

These two frequencies differ by less than the Rayleigh resolution, which is 
in our case $0.11$ \cd. This could indicate that the value we calculate for 
these frequencies might not be the real ones. However, the rapid amplitude 
variation present in the light curve, could hardly be explained by any 
other origin than the presence of two closely spaced frequencies, even if 
their real values are not exactly those given in Table~\ref{tabla freq nwser}.
Therefore we conclude that two close frequencies do actually exist, although 
their exact values might differ from those given here. In any case, 
\citet{loumos78} noted that the true frequencies always fall at the same 
midpoint as those found when the resolution is not high enough.

\begin{table}
\centering
\caption{Frequencies, amplitudes and phases obtained 
with the multi-parameter fitting code for NW~Ser  
for the four Str\"omgren bands. The SNR, 
the $\sigma$ of the final residual ($\sigma_{res}$) and
the total fraction 
of the variance removed from the signal (R) are also given.}
\begin{tabular}{ccccccc}
\hline
\hline
No. & Freq.  & Amp. & Phase   & SNR & $\sigma_{res}$ & R\\
 & \cd & mmag & 2$\pi$\,rad& & mmag & $\%$ \\
\hline
 & Filter y & & N=162 & \multicolumn{2}{c}{$\sigma_{init}=34.8 ~\mathrm{mmag}$}  &  \\
F1 & 1.190 & 38.1  &  0.84  & 25 & 20.8 &  \\
F2 & 1.119 & 21.8  &  0.88  &15  & 12.1& \\
F3 & 3.294 & 13.6  &  0.38 & 7  & 8.7&\\
F4 & 1.415 &  7.8  &  0.29  & 5  & 7.1& 96\\

 & Filter b & & N=162 & \multicolumn{2}{c}{$\sigma_{init}=34.9 ~\mathrm{mmag}$} &   \\
F1 & 1.196 & 36.6  &  0.85  & 25 & 22.0 &  \\
F2 & 1.127 & 26.9  &  0.86  &19 & 11.7& \\
F3 & 3.301 & 11.6  &  0.36 & 7  & 8.7&\\
F4 & 1.414 &  9.5  &  0.28  & 6  & 6.2& 97\\

 & Filter v & & N=162 & \multicolumn{2}{c}{$\sigma_{init}=35.1~ \mathrm{mmag}$}  &  \\
F1 & 1.197 & 36.4  &  0.86  & 28 & 22.6 &  \\
F2 & 1.126 & 27.7  &  0.88  & 22  & 12.1& \\
F3 & 3.304 & 11.3  &  0.30 & 7  & 8.5&\\
F4 & 1.412 & 10.5  &  0.36  & 7  & 6.2& 97\\

 & Filter u & & N=162 & \multicolumn{2}{c}{$\sigma_{init}=44.4~ \mathrm{mmag}$}  &  \\
F1 & 1.159 & 55.6  &  0.95  & 38 & 26.2 &  \\
F2 & 1.103 & 33.3  &  0.83  & 23  & 14.2& \\
F3 & 2.274 & 17.5  &  0.49 & 11  & 10.5&\\
F4 & 2.387 & 12.9  &  0.34  & 8  & 7.4& 97\\

\hline
\end{tabular}
\label{tabla freq nwser}
\end{table}

Residuals after prewhitening for these two frequencies have been analysed for 
additional periodic components. Frequencies $\mathrm{F3}=3.304$ \cd\ and 
$\mathrm{F4}=1.412$ \cd\ and their daily aliases have been detected in successive 
periodograms. These new frequencies have a much lower amplitude than F1 and F2, 
although they have a SNR greater than 4, as shown in 
Fig.~\ref{fig freq nwser}. Note that $\mathrm{F3}=3.304$ \cd\ could be 
a combination of the two high-amplitude frequencies F1 and F2, since
$\mathrm{F}3 \sim \mathrm{F}1+\mathrm{F}2+1$.
Finally, the four frequencies which minimise the 
residuals are those shown in Table~\ref{tabla freq nwser} for the four 
Str\"omgren bands. No other significant peaks appear in the periodogram after
 the final prewhitening.

It is important to note that the four frequencies obtained independently for 
 the different $vby$ filters are the same within the error boxes. For 
the $u$ filter we found the 1 \cd\ alias of the lower amplitude frequencies F3 and F4.
 This is due to the low SNR of the data at that wavelength. As seen in 
Table~\ref{tabla freq nwser},
 the percentage of the total fraction of the variance that is removed 
(R-values) for all filters is very high.

Note that the final frequencies can be contaminated by 1~\cd\ aliases, since the observations have been obtained at only one site.
Keeping this in mind, if we compare with the frequencies found by other authors, the frequency F2 could be a 1-day alias of the first frequency (f1\,$=2.11$ \cd) detected by \citet{aerts00} using Hipparcos data and F4 could be a 1-day alias of her second frequency (f2\,$=2.46$ \cd). In the same way, F1 could be the 
1-day alias of the frequency 2.17 \cd\ obtained by \citet{percy99}. The complexity of the Hipparcos spectral window due to the few (76) datapoints spanning $\sim$ 1000 days makes 
the detection of multiple short periods very difficult and uncertain. 
Nevertheless, to test our results we performed a re-analysis of the Hipparcos light curve with the non-linear least-square fitting code.
We have detected the two frequencies found by \citet{aerts00} as the best 2-frequency fitting. 
However, the best 3-frequency model is given by the frequencies 2.09, 1.43 and 3.30 \cd. 
The frequency F2 could be the 1-day alias of 2.09 \cd\ and the two latter are similar to F4 and F3 respectively.

In Table~\ref{tabla freq nwser} we list the frequencies resulting from the analysis
of the OSN data. The above considerations indicate that these
frequencies can be either real or in some cases their 1-day
aliases. The analysis of the Hipparcos data shows the presence of the
same frequencies or their aliases, confirming that NW Ser is a
multiperiodic variable, as first suggested by \citet{aerts00}, and that the pulsation modes 
remain stable at least
between the two epochs. To ascertain what
frequencies are real or the aliases, a more extended
photometric dataset with a better spectral window, either from
a ground-based multisite campaign or from space, would be needed.

\subsection{V1446~Aql}

 \begin{figure}
   \centering
   \includegraphics[width=7cm]{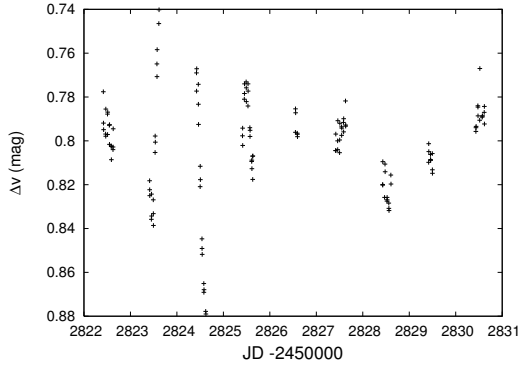}
   \caption{Light curve of V1446~Aql in the $v$ filter.}
         \label{fig curva v1446aql}
   \end{figure}

\begin{figure*}
    \centering
    \includegraphics[width=5.5cm]{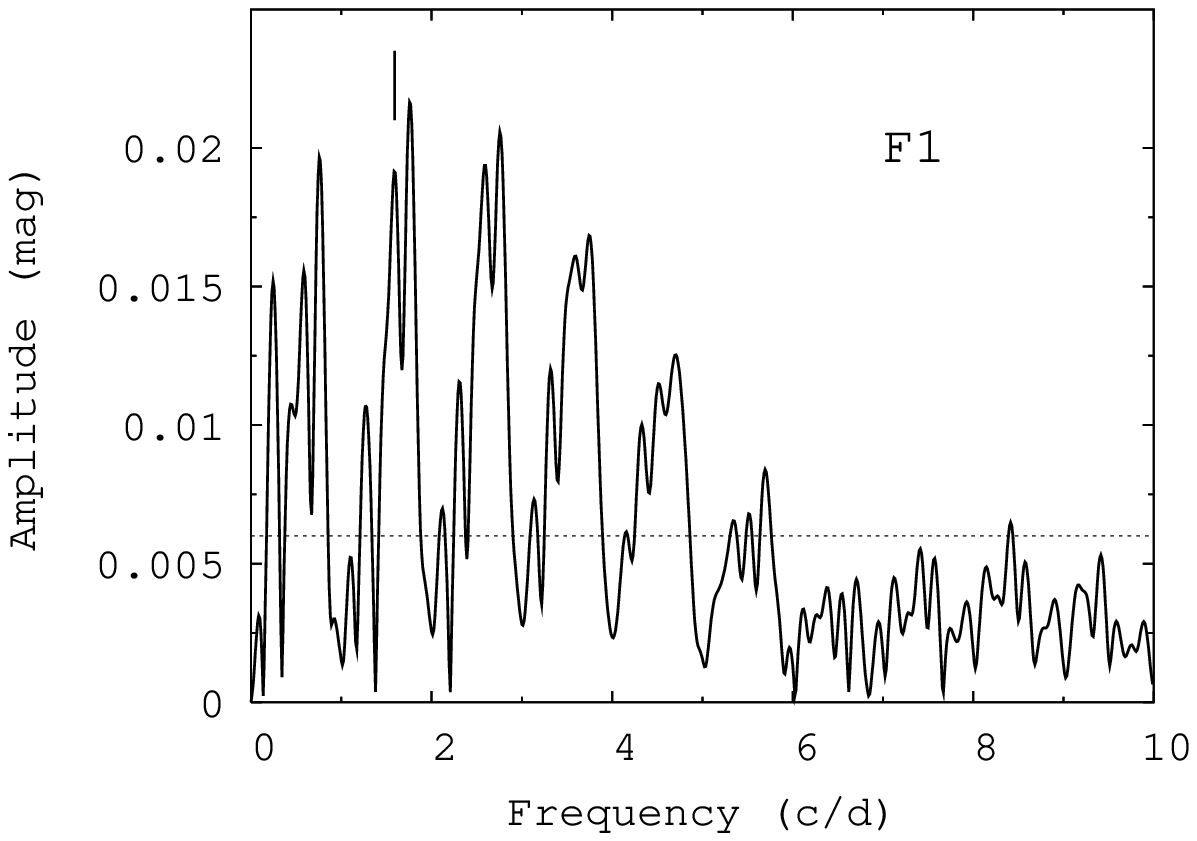}
    \includegraphics[width=5.5cm]{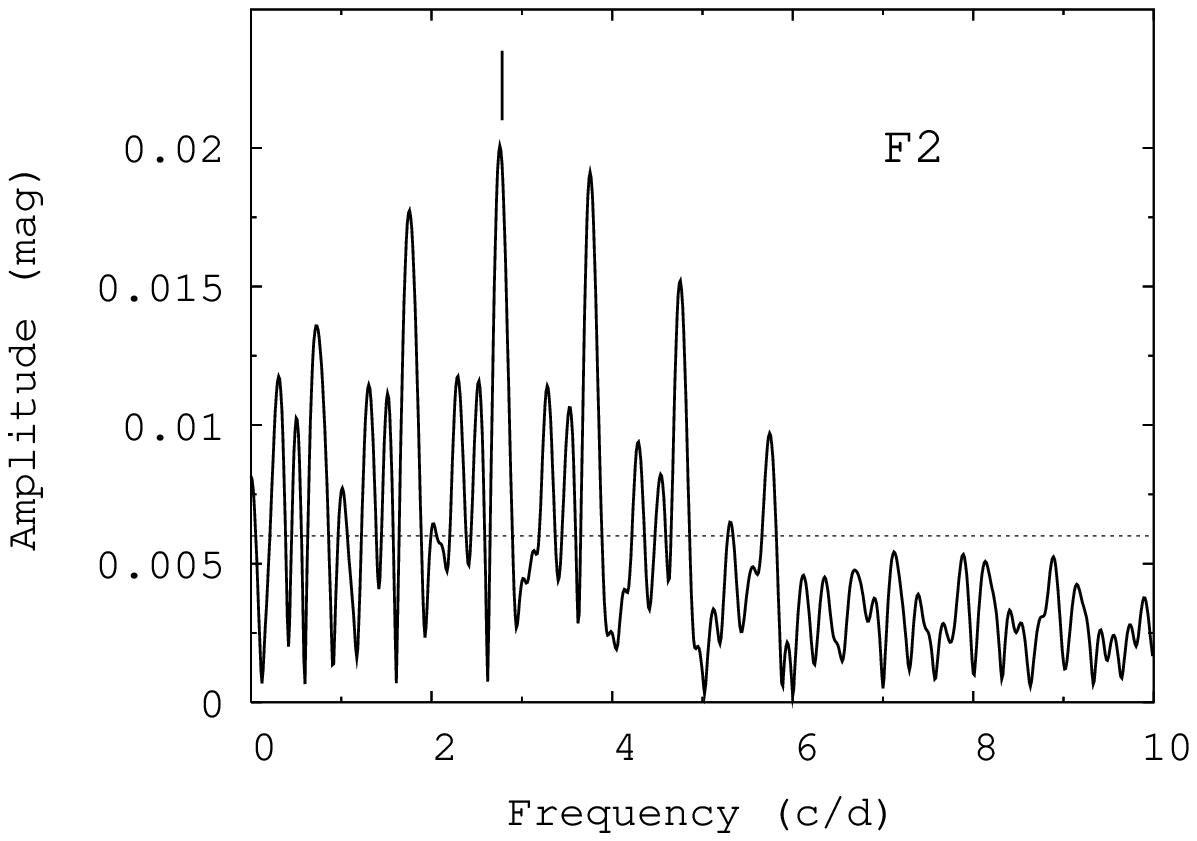}
    \includegraphics[width=5.5cm]{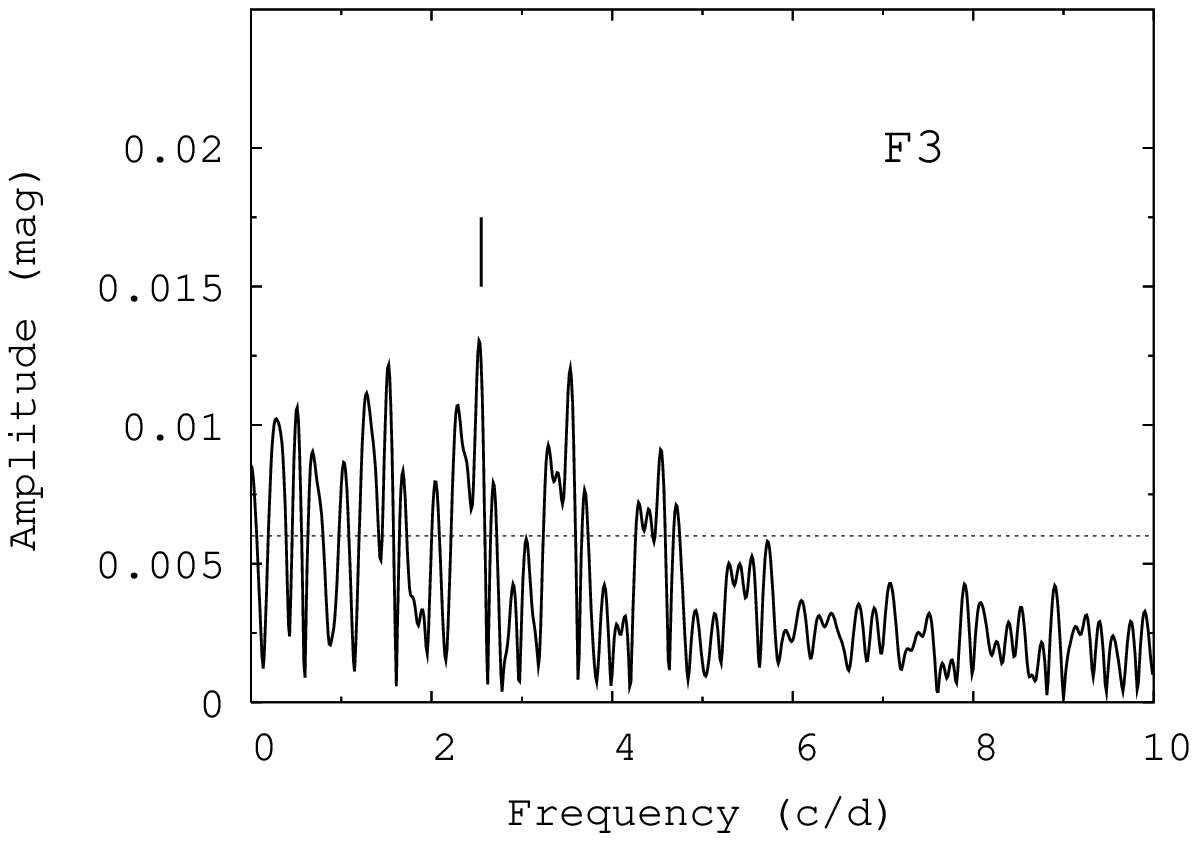}
    \includegraphics[width=5.5cm]{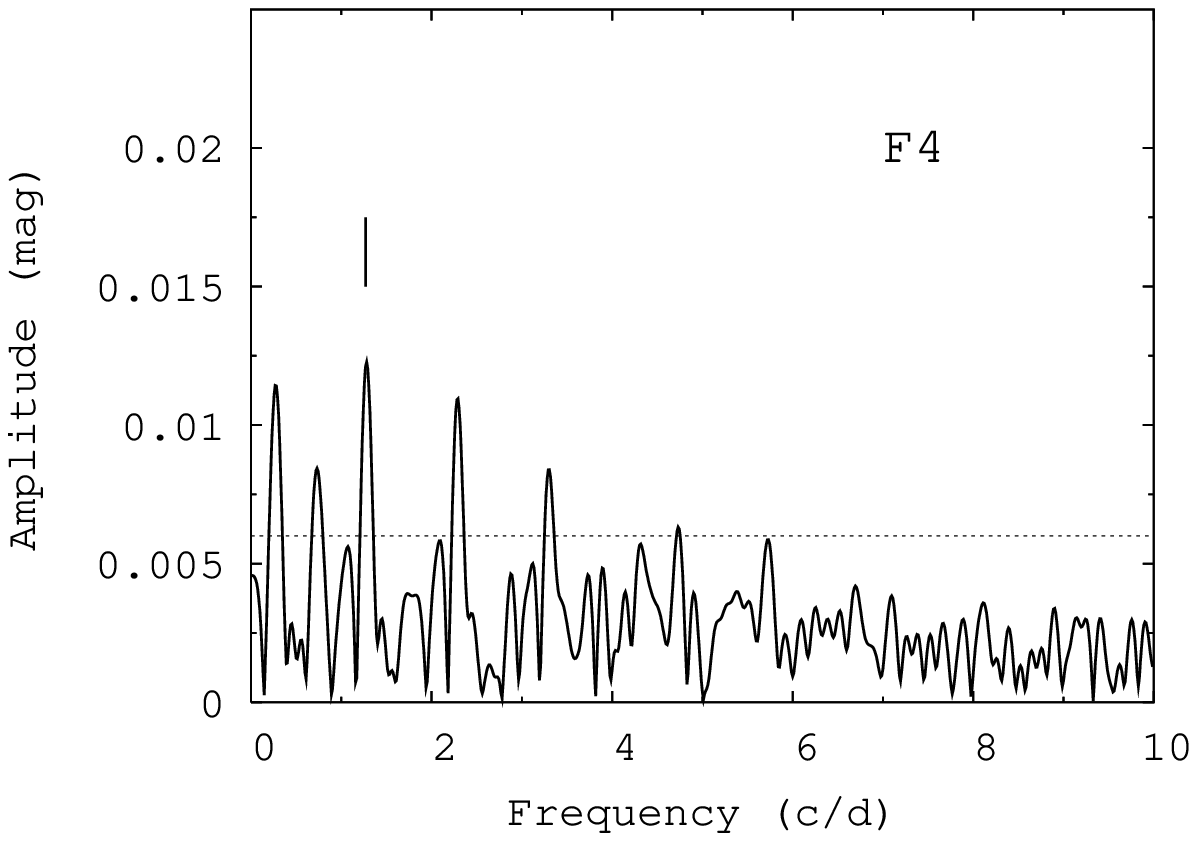}
    \includegraphics[width=5.5cm]{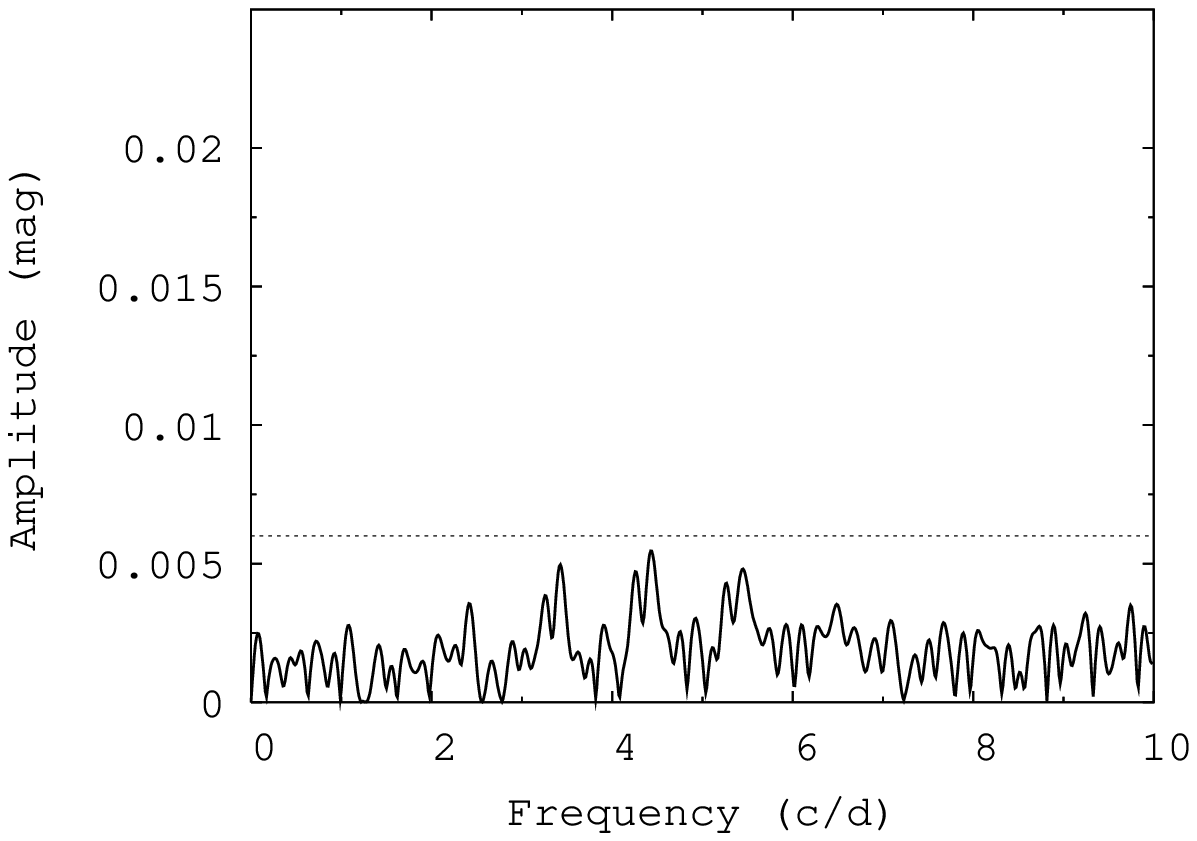}
    \includegraphics[width=5.5cm]{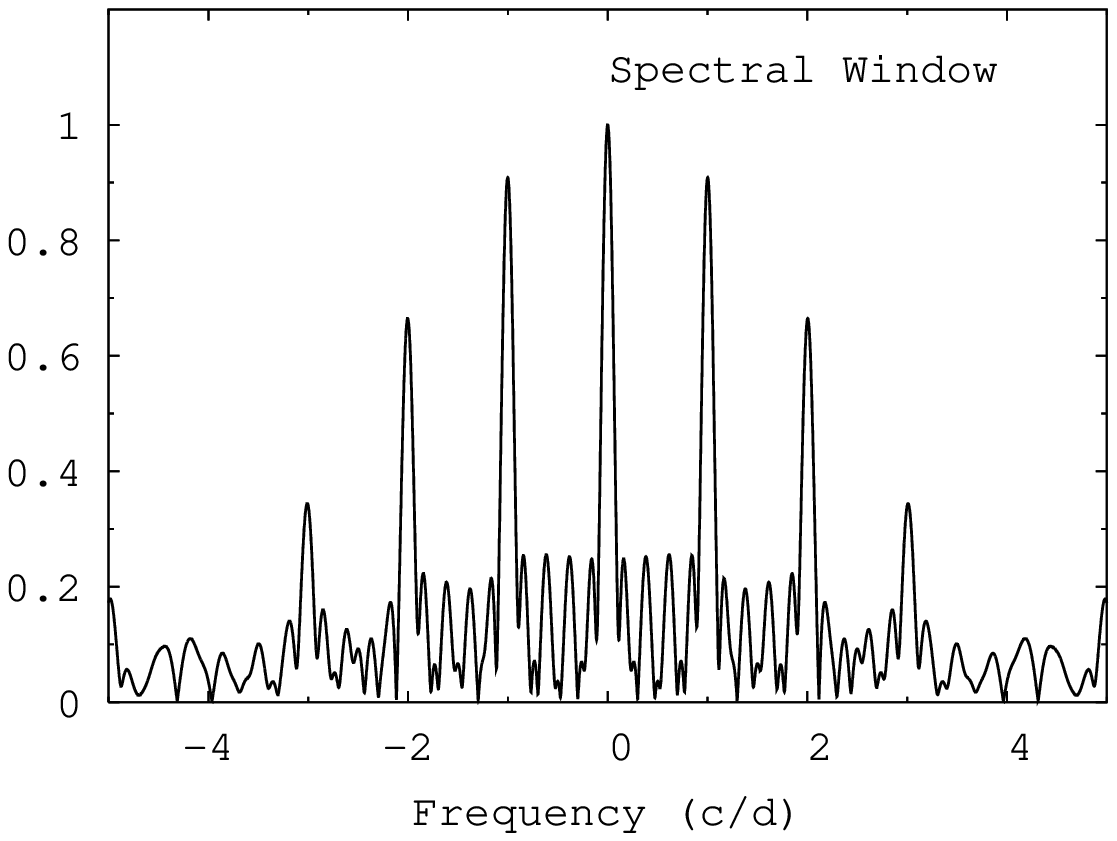}
    \caption{Successive periodograms for V1446~Aql for the $v$ filter. Symbols are
 the same as for Fig.~\ref{fig freq nwser}.}
\label{fig freq v1446aql}
\end{figure*}

\begin{table}
\centering
\caption{As for Table~\ref{tabla freq nwser}, but for V1446~Aql.}
\begin{tabular}{ccccccc}
\hline
\hline
No. & Freq.  & Amp. & Phase   & SNR & $\sigma_{res}$ & R\\
 & \cd & mmag & 2$\pi$\,rad& & mmag & $\%$ \\
\hline
 & Filter y & & N=136 & \multicolumn{2}{c}{$\sigma_{init}=23.89 ~ \mathrm{mmag}$} & \\
F1 & 1.631 & 21.1  &  0.76  & 17 & 19.6 &  \\
F2 & 2.792 & 18.8  &  0.21  &11  & 15.3& \\
F3 & 2.559 & 18.7  &  0.66 & 12  & 12.3&\\
F4 & 1.264 & 13.0  &  0.32  & 11  & 8.8& 87\\
 & Filter b & & N=136 & \multicolumn{2}{c}{$\sigma_{init}=24.85 ~ \mathrm{mmag}$}  & \\
F1 & 1.622 & 21.7  &  0.78  & 14 & 20.4 &  \\
F2 & 2.783 & 20.9  &  0.22  & 11  & 15.1& \\
F3 & 2.555 & 17.7  &  0.65 &  10  & 12.3&\\
F4 & 1.266 & 12.9  &  0.33  & 8  & 8.9& 87\\
 & Filter v & & N=136 & \multicolumn{2}{c}{$\sigma_{init}=24.84 ~ \mathrm{mmag}$}  & \\
F1 & 1.617 & 22.1  &  0.79  & 15 & 19.9 &  \\
F2 & 2.783 & 20.9  &  0.23  & 11  & 14.3& \\
F3 & 2.557 & 17.5  &  0.64 & 10  & 11.7&\\
F4 & 1.269 & 12.4  &  0.32  & 8  & 8.3& 89\\
 & Filter u & & N=136 & \multicolumn{2}{c}{$\sigma_{init}=29.10 ~ \mathrm{mmag}$}  & \\
F1 & 1.601 & 27.0  &  0.77  & 13 & 22.6 &  \\
F2 & 2.759 & 25.5  &  0.32  & 11  & 17.4& \\
F3 & 2.514 & 11.1  &  0.67 & 5  & 14.6&\\
F4 & 1.187 & 15.5  &  0.62  & 8  & 11.1& 85\\
\hline
\end{tabular}
\label{tabla freq v1446aql}
\end{table}
The light curve of V1446~Aql, presented in Fig.~\ref{fig curva v1446aql}, 
also shows rapid variation which is probably caused by the presence of a 
beating phenomenon between two close frequencies. As the frequencies found for 
the four Str\"omgren filters are basically the same, we only 
discuss here the detailed analysis for the $v$ filter, since it shows the highest 
SNR. Frequencies at 1.617 \cd\ and at 2.783 \cd\ and their daily aliases 
appear clearly in the periodogram. Note that these two frequencies have 
similar amplitude. The amplitude of the first frequency (1.617 \cd) fulfils the signal to 
noise criterion, as shown in Fig.~\ref{fig freq v1446aql}. After prewhitening for this frequency, the second frequency at 2.783 \cd, and its 
daily aliases, appear clearly. By prewhitening for these two frequencies, other 
components have been found. Either a frequency at 2.557 \cd\ or its 1-day alias at 
1.557 \cd\ can be considered as the third frequency, although 1.557 \cd\ is so 
close to F1 that the fitting algorithm does not converge and we obtain 
unrealistic values for this solution. Therefore we remove the 
frequency at $\mathrm{F3}=2.557$ and identify the fourth component 
$\mathrm{F4}=1.269$ \cd, or one of its aliases.

The final set of frequencies, displayed in Table~\ref{tabla freq v1446aql}, 
fulfils the signal to noise requirement in all the filters.
We stopped the frequency search at this point 
because new peaks did not fulfil the SNR criterion mentioned above.

As in the previous subsection, we list the frequencies corresponding to
the best fit to our OSN data. Due to the effects of data sampling
discussed above, some of the listed frequencies might be 1-day aliases
of the real ones.

\section{Theoretical modelling}
\begin{figure}
 \begin{center}
   \includegraphics[width=8cm]{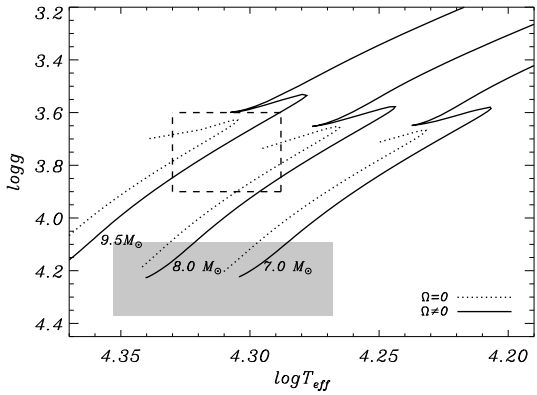}
   \caption{Hertzsprung-Russel diagram showing the two Be stars 
considered in this study and some relevant evolutionary tracks
	     selected for the modelling. The empty and shaded regions 
	     represent the corresponding error boxes for NW~Ser and
	     V1446~Aql respectively. 
	     Continuous and dotted lines represent evolutionary tracks 
	     of 7, 8 and $9.5\,\msol$ for rotating models 
	     and non-rotating models respectively.
}
   \label{fig:hr}
 \end{center}
\end{figure}

To determine whether the observed frequencies are predicted unstable or 
not by the current models of stellar pulsation we have made a preliminary 
theoretical study. The physical parameters of NW~Ser and V1446~Aql have been 
obtained from the analysis of high resolution spectra \citep{fremat06}, 
taking into account the effects of gravitational darkening and assuming both 
stars are rotating at 88\% of their critical break-up velocity 
\citep{fremat05_a}. The obtained values and associated errors are depicted in 
Fig.~\ref{fig:hr}. Note that both error boxes are located in the overlapping 
region of the $\beta$ Cephei and SPB instability strips \citep{pamyatnykh99}.

Evolutionary tracks computed with the numerical code CESAM \citep{morel97} are
shown in Fig.~\ref{fig:hr}.
First-order effects of rotation are taken into account in the
equilibrium models. To do so,
the equilibrium equations are modified in
the CESAM code in the manner described in \citet{kippenhahn90}.
The so-called \emph{pseudo}-rotating models include the spherically
averaged contribution of the centrifugal acceleration by
means of an effective gravity $g_{\mathrm{eff}}=g-{\cal A}_{c}(r)$,
where $g$ is the local gravity, $r$ is the radius,
and ${\cal A}_{c}(r)=\frac{2}{3}\,r\,\Omega^2(r)$ is the centrifugal
acceleration of matter elements. This spherically averaged component
of the centrifugal acceleration does not change the order of the hydrostatic
equilibrium equations. The models are assumed to rotate uniformly and to
conserve their total
angular momentum during evolution.

Considering the corresponding error boxes, 
masses in the range of 8.5--$9.5\,\msol$ are found to be representative of 
NW~Ser. Rotational velocities are in the range of $250$ to $270\,\kms$,
radii from $5.8$ to $7.9\,\rsol$ and ages around 20 Myr. In the same way, 
for V1446~Aql, a mass range of 7--$7.5\,\msol$ is found, with rotational 
velocities ranging from $238$ to $331\,\kms$, radii from $4$ to $4.5\,\rsol$ 
and an age that brings the star close to the ZAMS.

\begin{figure}
 \begin{center}
   \includegraphics[bb=50 50 410 280, clip, width=8cm]{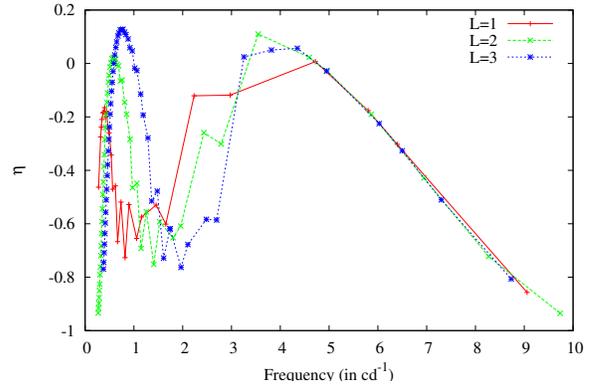}
   \caption{Growth rates diagram for NW~Ser: positive values indicate unstable 
 modes and viceversa.}
   \label{fig growth nwser}
 \end{center}
\end{figure}

\begin{figure}
 \begin{center}
   \includegraphics[bb=50 50 410 280, clip, width=8cm]{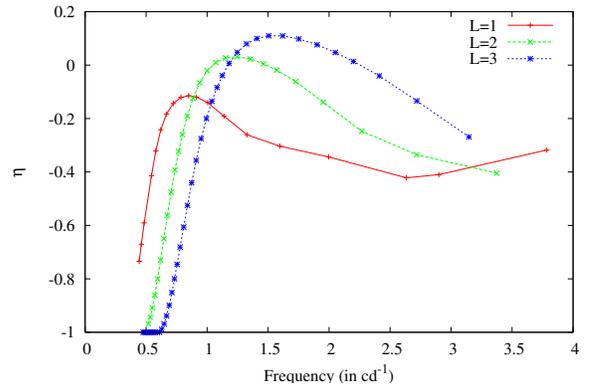}
   \caption{Growth rates diagram for V1446~Aql.}
   \label{fig growth v1446}
 \end{center}
\end{figure}

The instability analysis is then performed using the GraCo code 
\citep[see ][ for more details]{moya04}, which is based on the 
non-adiabatic equations derived by \citet{unno89}. 
For both stars, the $\ell=2$ and $\ell=3$ modes are predicted to be 
unstable. In the case of NW~Ser, the SPB and $\beta$ Cephei instability 
ranges are very close (Fig.~\ref{fig growth nwser}), and this suggests that NW~Ser is a 
good candidate to be a hybrid pulsator. 
This result is 
consistent with its location in the HR diagram region where the instability 
strips of SPB and $\beta$ Cephei overlap. In the case of V1446 Aql, only 
high-order g-modes with $\ell= 2$ and 3 are predicted to be unstable close to the
observed frequencies, as shown in Fig.~\ref{fig growth v1446}.

\begin{figure}
 \begin{center}
   \includegraphics[bb=100 30 590 690, clip,angle=-90,width=8cm]{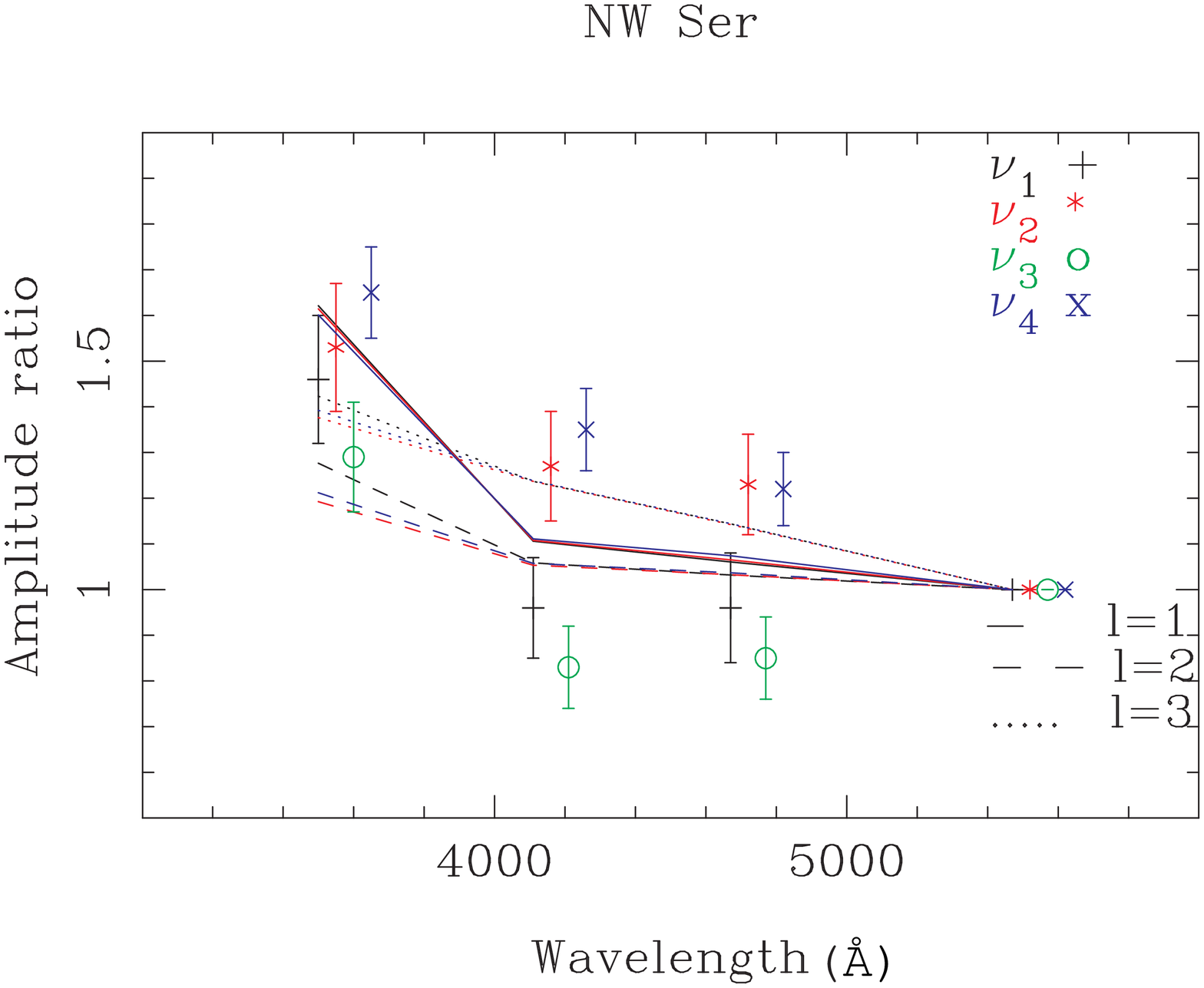}
   \caption{Observed amplitude ratios $A_X/A_y$ and their error boxes for the 
detected frequencies of NW~Ser. $A_X$ stands for any of the amplitudes in the 
$uvby$ Str\"omgren filters. The lines represent the theoretical 
non-adiabatic predictions of the stellar models explained in the text.}
   \label{fig colores nwser}
 \end{center}
\end{figure}

\begin{figure}
 \begin{center}
   \includegraphics[bb=100 30 590 690, clip,angle=-90,width=8cm]{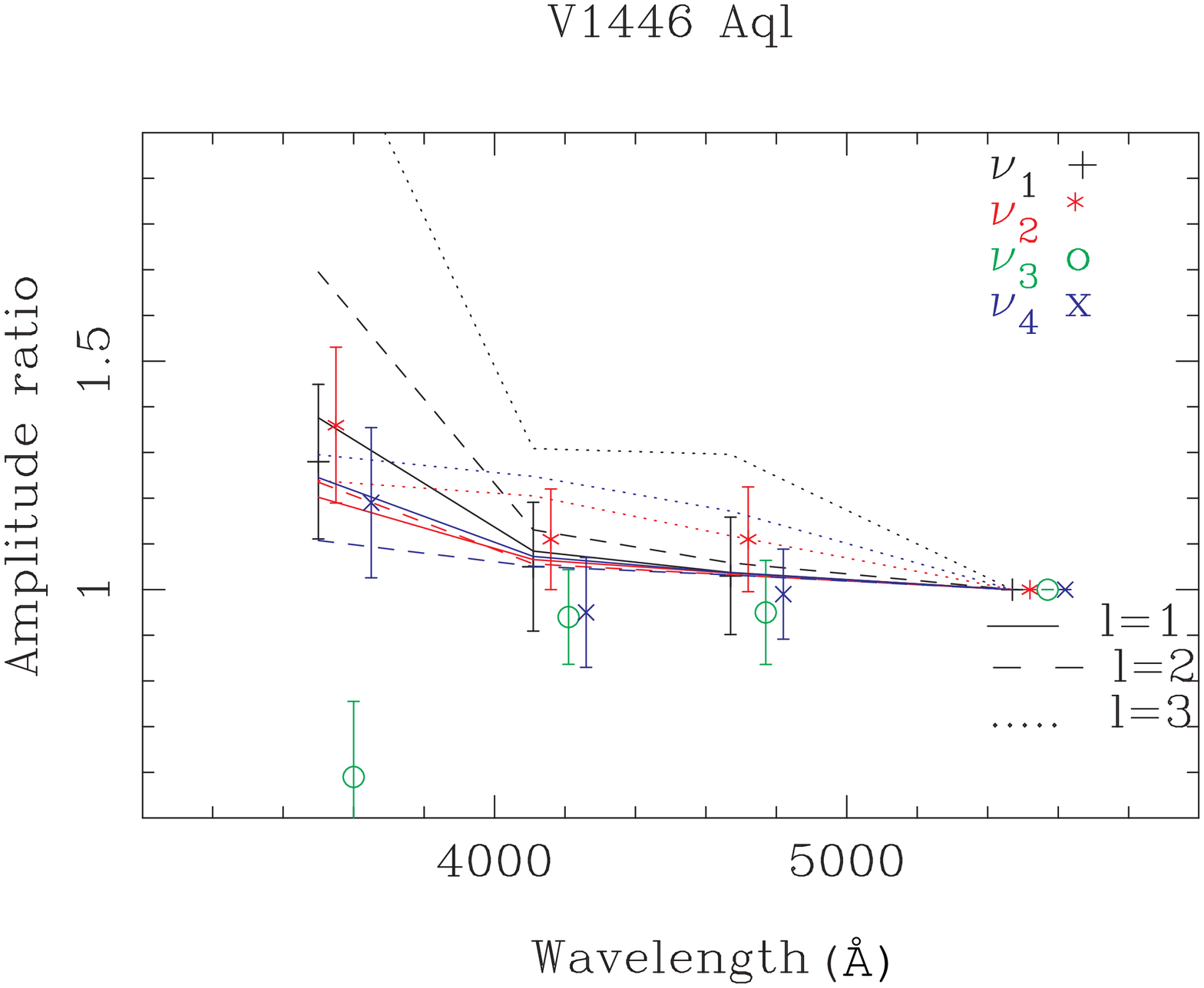}
   \caption{Same as Fig.~\ref{fig colores nwser}, but for V1446~Aql.
}
   \label{fig colores v1446aql}
 \end{center}
\end{figure}

In order to have an idea of what spherical degree $\ell$ is excited in these stars 
we have applied the photometric mode identification as described in \citet{watson88} and
 used for $\delta$ Scuti stars in \citet{garrido00}. We have computed the 
amplitude ratios for each pulsational frequency and we have compared them with 
those predicted for stellar models using different $\ell$-degrees. In 
Figs.~\ref{fig colores nwser} and~\ref{fig colores v1446aql} we show the 
amplitude ratios for both stars. In the case of NW~Ser, pulsation modes of 
degree $\ell=3$ appear to be more probable for frequencies F2 and F4, whereas for
 F1 and F3 the more probable degrees seem to be $\ell=1$ and 2. For 
V1446~Aql all detected frequencies seem to be associated with dipoles and 
quadrupoles. The $u$ amplitude for frequency F3 is very low and probably due to the 
noise of the data at this wavelength.

At this point only spectroscopic observations can supply an unambiguous mode 
identification for these rapidly rotating objects.

\section{Conclusions}

Accurate photometric light curves of the Be stars NW~Ser and V1446~Aql
obtained during a preparatory program for the COROT space mission
have been analysed. 
Our spectral analysis reveals the presence of four statistically
significant frequencies in each star.
Together with the multiperiodicity of 
HD 163\,868 and $\zeta$-Oph recently observed from space \citep{walker05a,walker05b}, the present 
results point towards the interpretation of short-period
variability of Be stars in terms of $nrp$, as suggested 
by spectroscopic observations obtained in the last decade. In this
framework, a theoretical instability analysis of the observed
frequencies has been performed. 
Recent studies \citep{saio07,Dziembowski07} show that, for very rapid rotators, such as Be
stars, 
the effect of rotation may also affect the instability of the modes. In
future
studies these effects should then be taken into account and compared with
the present results.
It is found that NW~Ser 
could be an hybrid pulsator, since 
the ranges of unstable g-modes (SPB-like) and p-modes ($\beta$ Cephei-like) are very close. 
This would thus imply
that NW~Ser is a good candidate of the very few stars detected so far showing such 
pulsational characteristics. In the case of V1446~Aql, only frequencies 
corresponding to predicted g-modes have been detected.

Longer timebase spectroscopic and photometric observations are 
thus required in order to confirm these results.


\begin{acknowledgements}
This research is based on data obtained at the Observatorio de Sierra Nevada,
 which is operated by the CSIC through the Instituto de Astrof\'{\i}sica de 
Andaluc\'{\i}a. The work of J.G-S. is supported by a FPU grant from the 
Spanish ``Ministerio de Educacion y Ciencia''.
J.F. and J.S. acknowledge financial support from the program 
ESP~2004-03855-C03. 
J.C.S acknowledges support by the Instituto de Astrof\'{\i}sica de Andaluc\'{\i}a
by and I3P contract financed by the European Social Fund and from the Spanish 
Plan Nacional del Espacio under project ESP2004-03855-C03-C01.
\end{acknowledgements}

\bibliographystyle{aa}
\bibliography{7414.bib}
\end{document}